\documentclass[iicol,pdflatex,sn-aps]{sn-jnl}
\usepackage[T1]{fontenc}
\usepackage{amsmath}
\usepackage{amsfonts}
\usepackage{amssymb}
\usepackage{graphicx}
\usepackage{bm}
\usepackage{isotope}
\usepackage{siunitx}
\usepackage{float}
\usepackage{multirow}
\usepackage{booktabs} 
\usepackage{hyperref}
\usepackage{lmodern}

\begin{document}

\newcommand{\tke}{T\!K\!E}
\newcommand{\avg}[1]{\ensuremath{\langle #1 \rangle}}
\newcommand{\labr}{$\mathrm{LaBr}_3$(Ce)}
\newcommand{\hf}{\ensuremath{T_{1/2}}}
\newcommand{\tss}[1]{\textsuperscript{#1}} 

\title{Characterization of isomers produced by the spontaneous fission of \texorpdfstring{\tss{252}{Cf}}{252Cf} with the VESPA setup}

\author[1,2]{V.~Piau}
\author[3]{A.~G\"o\"ok}
\author*[4]{S.~Oberstedt}\email{stephan.oberstedt@ec.europa.eu}
\author[5]{A.~Oberstedt}
\author[1]{A.~Chebboubi}
\author[1]{O.~Litaize}
\author[4]{M.~Vidali}
\affil[1]{CEA/DES/IRESNE/DER/SPRC/LEPh, Cadarache, 13108 Saint-Paul lez Durance, France}
\affil[2]{Université Paris-Saclay, CNRS/IN2P3, IJCLab, 91405 Orsay, France}
\affil[3]{Department of Physics and Astronomy, Uppsala University, Box 516, 751 20 Uppsala, Sweden}
\affil[4]{European Commission, Joint Research Centre, Directorate for Nuclear Safety and Security, 2440 Geel, Belgium}
\affil[5]{Extreme Light Infrastructure - Nuclear Physics (ELI-NP), Horia Hulubei National Institute for Physics and Nuclear Engineering (IFIN-HH), 077125 Bucharest-Magurele, Romania}

\abstract{
Isomers produced by spontaneous fission of ${}^{252}$Cf have been measured with the VESPA setup, composed of LaBr$_3$(Ce) detectors for fast $\gamma$-ray spectroscopy and an ionization chamber for detecting fission fragments. Identification of the isomers was derived from fission fragment-$\gamma$-$\gamma$ coincidences.
This paper presents the half-life of 34 isomeric states measured with this setup, from less than the nanosecond up to tens of microseconds. Two of these isomers are reported for the first time, in ${}^{108}$Tc and ${}^{147}$Ce.
In addition to this half-life analysis, the isomers are used to develop and test a nuclear charge calibration of the ionization chamber.
}


\maketitle

\section{Introduction}

At EC-JRC Geel, the VESPA setup~\cite{Travar2021}, made of a position-sensitive double Frisch-grid ionization chamber surrounded by fast \labr{} scintillation detectors, is dedicated to multi-parameter measurements on spontaneous fission.
It was recently used to study the prompt fission $\gamma$-ray emission correlated with fission fragments characteristics (mass and kinetic energy) for \isotope[252]{Cf}(sf)~\cite{Travar2021,Piau2023}.
These studies focused on the prompt peak, i.e., $\gamma$-rays emitted within the first nanosecond after the scission of the fissioning nucleus. This prompt peak alone amounts to 90\% of all $\gamma$-rays emitted by the fission fragments~\cite{Skarsvag1975}. The remaining $\gamma$-rays are emitted from isomeric states of fission fragments, at a longer time scale (up to several \si{\micro\second}). 

The analysis of these isomeric states has several interests. First, isomers are a powerful probe for understanding the fission process. In particular, the angular momentum generation of the fission fragments, that is not yet fully understood, can be obtained from isomeric yield ratios, see e.g.~\cite{Chebboubi2017,Gjestvang2023}.
In that regards, codes modeling the prompt de-excitation of fission fragments, like FIFRELIN~\cite{Litaize2015}, are a key ingredient to extract the angular momentum from isomeric yield ratio measurements. Therefore, the accurate knowledge of fission fragment level schemes, including the half-life of any isomeric state, is required in order to assess with high reliability the derived fission fragment angular momentum.
Second, the measurement of isomer half-lives has proven to be a fruitful source of information for nuclear structure studies. In that respect, spontaneous fission is a convenient reaction, that produces neutron-rich nuclei far from stability without the need for neutron or radioactive ion beams.

Hence, isomeric states produced in the spontaneous fission of \isotope[252]{Cf} have been studied using the VESPA setup. This setup benefits from excellent timing characteristics, from both a state-of-the-art ionization chamber and fast scintillation detectors. The present study was based on fission-tagged $\gamma$-$\gamma$ coincidences between five almost identical $\gamma$-ray detectors.

The present paper is organized as following. Section~\ref{sec:setup} is dedicated to the detailed description of the VESPA experimental setup, and the characteristics of its detectors. Section~\ref{sec:analysis} presents the data analysis procedure used to identify and analyse the fission fragments isomers. In Section~\ref{sec:halflife}, we present the half-lives of the isomers measured in this work, ranging from nanosecond to several microseconds. In particular, we report two new isomeric states, in \isotope[108]{Tc} and \isotope[147]{Ce}. 
Section~\ref{sec:chargeIC} shows how we made the most of this experimental data to propose a detailed calibration procedure of the VESPA ionization chamber with respect to the nuclear charge of the fission fragments. Finally, Section~\ref{sec:conclu} summarizes the main results of this work.

\section{Experiment}

\subsection{Setup} \label{sec:setup}

\begin{figure}[ht]
    \centering
    \includegraphics[width=220pt]{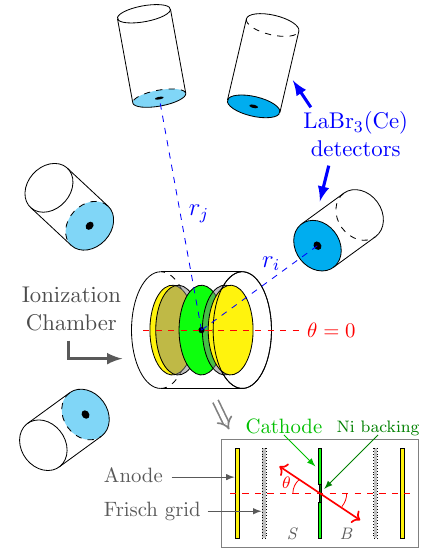}
    \caption{Schematic representation of the VESPA setup for isomers studies, consisting in five $2''\times2''$ \labr{} detectors placed around a twin Frisch-grid ionization chamber. \textit{B} (resp. \textit{S}) indicates the backing side (resp. source side) of the IC.}
    \label{fig:vespa}
\end{figure}

The experiment took place at the EC-JRC Geel (Belgium) with the VESPA setup for fission studies~\cite{Travar2021}. A source of \isotope[252]{Cf} undergoing spontaneous fission was placed inside of a twin Frisch-grid ionization chamber (IC)~\cite{Gook2016} filled with methane, which serves as a fission detector. The Cf source is deposited on a nickel backing, with a thickness of \qty{250}{\nm}, creating a small asymmetry between the 'backing side' and the 'source side', as fragments will lose some additional kinetic energy passing through this backing. 
The $\gamma$-ray spectroscopy was performed by five $2''\times2''$ (\qty{51}{\mm} diameter $\times$ \qty{51}{\mm} length) cylindrical \labr{} scintillation detectors surrounding the ionization chamber. These detectors were placed perpendicular to the target normal, at distances $r$ between 13.1 and \qty{14.6}{\cm} from the \isotope[252]{Cf}. A schematic view of this setup is depicted in Figure~\ref{fig:vespa}, and the position of the \labr{} detectors are reported in Table~\ref{tab:labr}.

As described in~\cite{Travar2021}, the \labr{} detectors were energy-calibrated from \qty{80}{\keV} to \qty{9}{\MeV} using standard sources. The energy resolution of the detectors has also been measured from this calibration. For $E_\gamma=\qty{661.7}{\keV}$, the relative Full Width at Half Maximum (FWHM) is between \qty{2.7}{\percent} and \qty{3}{\percent}, as indicated in Table~\ref{tab:labr}.

\begin{table}[ht]
    \setlength{\tabcolsep}{8pt}
    \centering
    \caption{Characteristics of the \labr{} detectors used in the VESPA setup. The position of the detectors is defined by the distance $r$ between the source (at the center of the IC cathode) and the detectors front face, and the azimuthal angle $\varphi$. The relative energy resolution (FWHM) is also given for the \isotope[137]{Cs} full energy peak.}
    \label{tab:labr}
    \begin{tabular}{@{}ccc@{\hskip 6ex}c@{}}
        \toprule
        \labr & \multicolumn{2}{c}{Position} & Resolution\\
        serial no. & Distance $r$ & Angle $\varphi$ & (\qty{662}{\keV}) \\
        \midrule
        Q489 & \qty{13.98}{\cm} & \ang{139} & 2.81\%\\
        Q491 & \qty{13.12}{\cm} & \ang{41} & 2.97\%\\
        5414 & \qty{13.33}{\cm} & \ang{91} & 2.70\%\\
        5415 & \qty{13.43}{\cm} & \ang{-2}& 2.67\%\\
        5416 & \qty{14.58}{\cm} & \ang{180}& 2.76\%\\
        \bottomrule
    \end{tabular}
\end{table}

\begin{figure}[ht]
    \centering
    \includegraphics[width=0.99\linewidth]{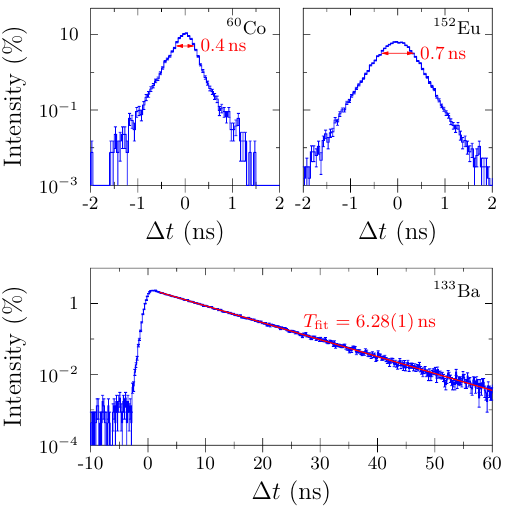}
    \caption{Coincident $\gamma$-$\gamma$ time spectra between any \labr{} combination from standard sources : \isotope[60]{Co} (\qty{1173}{\keV}, \qty{1332}{\keV}), \isotope[152]{Eu} (\qty{344}{\keV}, \qty{799}{\keV}) and \isotope[133]{Ba} (\qty{80}{\keV}, \qty{356}{\keV}).}
    \label{fig:calib}
\end{figure}

In addition to their good energy resolution, the \labr{} detectors also have excellent timing properties.
$\gamma$-$\gamma$ coincidence measurements were performed on various calibration sources, as depicted in Figure~\ref{fig:calib}. The time resolution (FWHM) associated to these coincidence spectra is about \qty{0.4}{\ns} for the \isotope[60]{Co} (\qty{1173}{\keV} and \qty{1332}{\keV}) and worsens at lower energies, as it can be seen for the \isotope[152]{Eu} coincidence (\qty{344}{\keV} and \qty{799}{\keV}). The \isotope[133]{Ba} time spectrum (\qty{80}{\keV}/\qty{356}{\keV} coincidence) demonstrates how the good time resolution of the detectors makes them well suited for half-life determination at the nanosecond scale.

\begin{figure}[ht]
    \centering
    \includegraphics[width=0.99\linewidth]{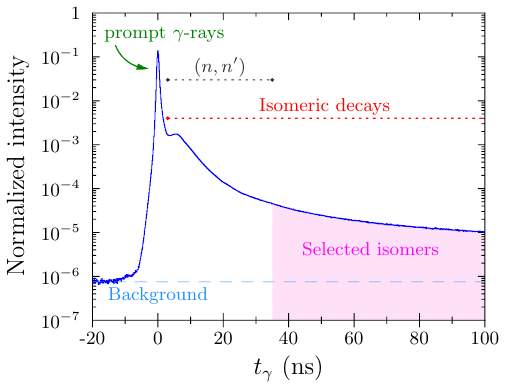}
    \caption{Time distribution of $\gamma$-rays detected in a \labr{} detector with respect to fission trigger from the IC.}
    \label{fig:tgamma}
\end{figure}

Along with these $\gamma$-ray detectors, the setup contains an ionization chamber similar to the one described in~\cite{Gook2016}, which serves two purposes.
First, it gives the fission trigger, i.e., the time stamp corresponding to the detection of a fission event. This trigger, combined with the timing information from the scintillators, is used to sort the $\gamma$ events relative to fission, which enables us to distinguish prompt transitions from isomeric level decays. As can be observed in Figure~\ref{fig:tgamma}, there are three main sources of $\gamma$-rays following fission : the $\gamma$-rays coming from the decay of the isomers, but also the prompt $\gamma$-rays (prompt peak), and neutrons. This last contribution comes from the inelastic scattering of prompt fission neutrons with the detectors or IC materials, mostly Al and Fe, generating spurious $\gamma$-rays in the data. 
The width of the prompt peak in Figure~\ref{fig:tgamma} is about \qty{0.6}{\ns} (FWHM), hence demonstrating the good timing resolution of the ionization chamber, as expected from this kind of detector~\cite{Budtz1987}.

The second purpose of the IC is the determination of the mass and energy of the Fission Fragments (FFs). These characteristics are estimated using the double kinetic energy (2E) method (see e.g.~\cite{Gook2018}). This method is based on mass and linear momentum conservation laws, and the simultaneous measurement of post-neutron FF kinetic energies. Neutron emission is accounted for in an iterative procedure, the average multiplicity $\bar{\nu}(A)$ taken from~\cite{Gook2014}, and assuming isotropic neutron emission in the center of mass frame. These simplifying assumptions are responsible for a smearing of the mass distributions obtained by this method~\cite{Gavron1974,Terrell1962}. As a consequence, the post-neutron mass resolution of the IC is about~\qty{5}{u} (FWHM).
The angle $\theta$ between the fission axis and the cathode's normal was determined from the electron drift time in the IC~\cite{Gook2016}.
Then, FFs emitted at grazing angles, that is $\cos\theta<0.5$ ($\theta<\ang{60}$), were rejected from the analysis. Indeed, such fragments may go through large energy loss, which would lead to incorrect energy and mass characterization.

Finally, the data obtained in this work have been acquired for about 3500 hours (effectively). The corresponding data amounts to \num{8.7e9} total fission events, from which \num{4.8e9} were selected from the cuts described above and analyzed in this work.

\subsection{Data analysis} \label{sec:analysis}

The isomers were selected by means of FF-$\gamma$-$\gamma$ coincidences, that is, a fission followed by two $\gamma$ rays detected in coincidence in \labr{} detectors. In this work, the coincidence window between two $\gamma$-rays has been set to $\pm\qty{2}{\ns}$. This short window is consistent with the timing properties of \labr{} scintillators, as discussed previously.

In order to remove the background caused by prompt $\gamma$-rays and neutrons and to keep only the $\gamma$-rays from isomers, a time cut has been applied to the analysed data.
From the neutrons time of flight from the chamber to the \labr{} detectors (see Table~\ref{tab:labr} for the distances), it has been calculated that the contribution from neutrons having a kinetic energy $E_n\gtrsim\qty{100}{\keV}$ is gated out by selecting only $\gamma$-rays detected more than \qty{35}{\ns} after fission.
Hence, in the following \textit{late emission $\gamma$-rays} will refer to $\gamma$-rays emitted after this \qty{35}{\ns} threshold.
It is noteworthy that for such times the fission fragments have already been stopped inside the chamber, hence no Doppler broadening is expected for the late emission $\gamma$-rays.

\begin{figure}[ht]
    \centering
    \includegraphics[width=0.99\linewidth]{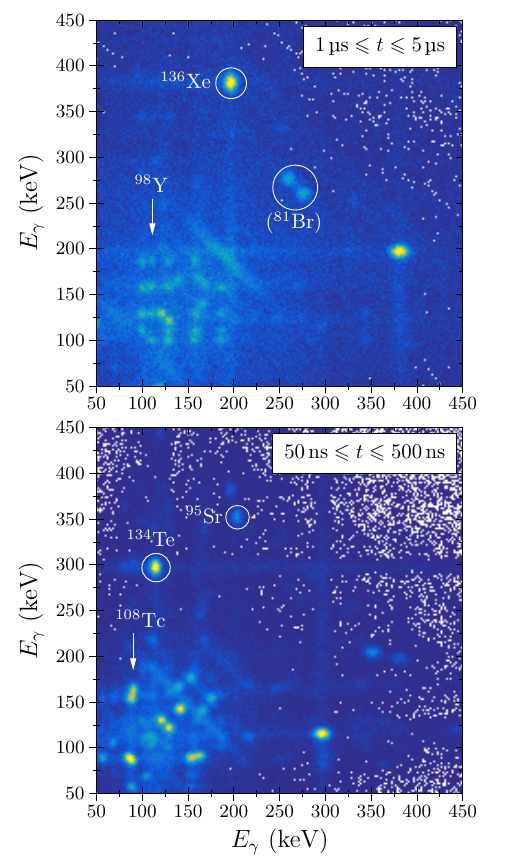}
    \caption{$\gamma$-$\gamma$ coincidence matrix obtained with VESPA at different time cuts relative to fission, between 50 and \qty{450}{\keV}.}
    \label{fig:ggcoinc}
\end{figure}

The identification of the isomers found in the VESPA coincidence data was performed using the NuDat database~\cite{Nudat,Nudat3}, based on ENSDF Adopted Levels and Gamma files. This analysis was based on the energy and time information of the detected gamma-rays.
For instance, $\gamma$-$\gamma$ matrices with different time cuts are shown in Figure~\ref{fig:ggcoinc}. In these matrices, several spots are visible, corresponding to abundant $\gamma$-$\gamma$ coincidences, from isomeric $\gamma$ cascades that could easily be identified.

\begin{figure}
    \centering
    \includegraphics[width=0.99\linewidth]{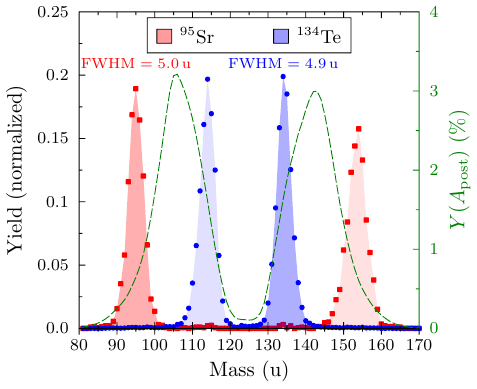}
    \caption{Mass distributions associated of \isotope[95]{Sr} (red) and \isotope[134]{Te} (blue) isomers from the VESPA ionization chamber. The post-neutron mass distribution of fission fragments in the spontaneous fission of \isotope[252]{Cf} is represented in green.}
    \label{fig:mass}
\end{figure}

The identification could be confirmed by studying the post-neutron mass distribution of the corresponding fission fragments, obtained from the IC.
Indeed, when gating on a $\gamma-\gamma$ coincidence corresponding to a particular isomer, the resulting mass distribution consists in a double gaussian. This corresponds to the emitting fragment and its complementary ones, smeared by the mass resolution of the IC, see Figure~\ref{fig:mass}.
On the contrary, no distinct peak comes from coincidences that do not originate from a specific fragment. This is for instance the case of the $\gamma$-$\gamma$ coincidence from $\isotope[81]{Br}$ that comes from the \labr{} scintillator, and can be excited up to an isomeric state ($E^*=\qty{536}{\keV}$ and $\hf=\qty{34.6}{\us}$) by the inelastic scattering of prompt fission neutrons in the crystals. The mass distribution associated to such spurious coincidence follows the $Y(A_\mathrm{post})$ distribution represented in Fig.~\ref{fig:mass}.

The half-life associated to these identified isomers was extracted by fitting their time distribution with respect to fission, with an exponential decay function. 
This commonly-used method is well suited to simple cases, where an isomeric state decays by emitting at least two gamma-rays in cascade, with a half-life larger than a few nanoseconds (due to the \qty{35}{\ns} threshold). It is also necessary that the nucleus studied by this method does not contain consecutive isomers having similar half-lives (e.g. \isotope[97]{Sr}), that would bias the measurements. 

To study more complicated level schemes having consecutive isomeric states, i.e., an isomer populated by another higher-lying isomeric state, we developed a complementary analysis procedure.
In this second approach, we looked at FF-$\gamma$ coincidences of previously identified isomers, gated on fragment mass, $\gamma$-ray energy and time. 
The isomeric state on which these gates are applied is referred to as the \textit{main} isomer. Then, all $\gamma$-rays from the same fission event and detected before this gated transition are recorded.
This method can highlight a \textit{populating} isomer, that is an isomeric state whose decay feeds the main isomer.
It could also be used to pinpoint \textit{intermediate} isomers, that are underlying isomeric states at one stage of the $\gamma$ cascade, populated by the main isomer. In particular, the method can be used to extract very short-lived isomers (a few ns) fed by longer isomers. 

\begin{figure}[ht]
    \centering
    \includegraphics[width=0.99\linewidth]{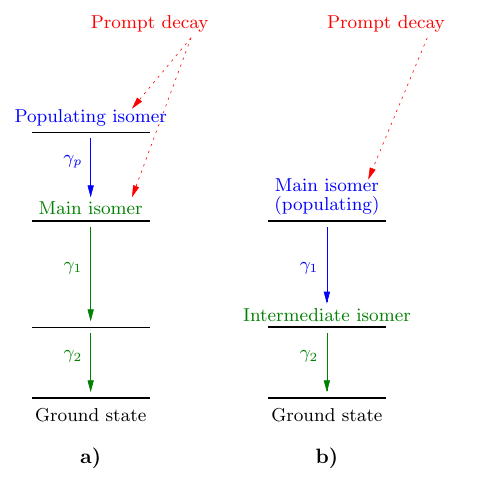}
    \caption{Schematic principle of the 'multiple isomers analysis' analysis for consecutive isomeric states. The main isomer, to which gates are applied, is represented in green, while populating $\gamma$-rays from prompt and isomeric decays are represented in red and blue, respectively.}
    \label{fig:delayed}
\end{figure}

These two cases are depicted in Figure~\ref{fig:delayed}. The half-life of the two consecutive isomers in play can be estimated by constructing the time distribution of the populating $\gamma$-ray, and the distribution of the time difference $\Delta{}t$ between these two $\gamma$-rays, for the populating and underlying isomeric states, respectively. In the following, this method will be referred to as the 'multiple isomers analysis'.

\begin{figure}[ht]
    \centering
    \includegraphics[width=\columnwidth]{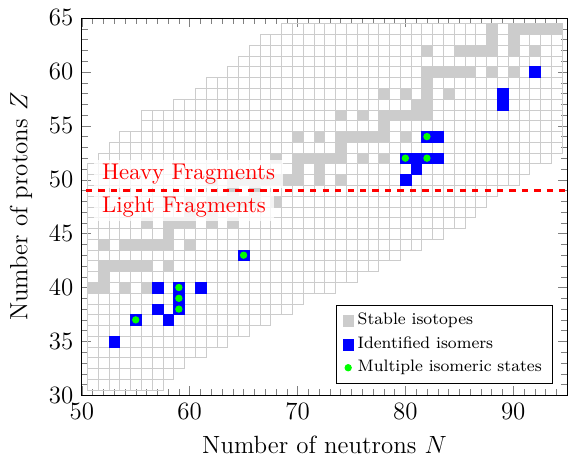}
    \caption{Chart of the identified nuclei having one or more isomeric state(s) whose half-life was measured with VESPA.}
    \label{fig:chart}
\end{figure}

Through this analysis of the VESPA data, obtained from the spontaneous fission of \isotope[252]{Cf}, isomeric states in 21 different nuclei were identified, from $A=88$ to $A=152$. These nuclei are depicted in Figure~\ref{fig:chart}. It is noticeable in this figure that we measured isomers from various fission fragments.
As presented before, additional isomeric states from these nuclei could be accessed from the multiple isomers analysis. In the end, the half-life of 34 isomeric states have been measured.

The half-lives measured in this work for all these isomers are reported in Table~\ref{tab:results}. In this table, the total uncertainties are given, corresponding to both statistical and systematic (syst1) errors.
These systematic uncertainties were estimated by varying the range of the fit, as well as the binning of the time histogram, as we could observe that the extracted half-lives were sensitive to these parameters.

\section{Results and discussion}

\subsection{Half-life extraction of the identified isomers} \label{sec:halflife}

\begin{table*}[p]
    \renewcommand{\arraystretch}{1.14}
    \centering
    \caption{Half-lives of the isomers measured in this work ($\hf^{\text{exp.}}$) compared with literature ($\hf^{\text{lit.}}$) and RIPL3-2023 database~\cite{RIPL} ($\hf^{\text{RIPL3}}$). In each nucleus, the "main isomer" is marked with an asterisk(*).}
    \label{tab:results}
    \begin{tabular}{@{}ccccc@{\hspace{5ex}}c@{}}
        \toprule
        & $E^*$ (MeV) & $\hf^{\text{exp.}}$ (ns) & $\hf^{\text{lit.}}$ (ns) & Ref. & $\hf^{\text{RIPL3}}$ (ns) \\ \midrule
        \isotope[88]{Br}  & 0.270* & 4500 (400) & 5500 (100) & \cite{Rzaca2018} & 5300 \\ \cmidrule{1-2}
        \isotope[92]{Rb}  & 0.284* & 54.4 (27) & 54 (3) & \cite{Urban2012} & 54\\
                          & 0.142\;\; & 0.82 (4) & 0.75 (3) & \cite{ND92} & 0.75 \\ \cmidrule{1-2}
        \isotope[95]{Rb}  & 0.835* & 101 (24) & 94 (7) & \cite{Simpson2010} & -- \\\cmidrule{1-2}
        \isotope[95]{Sr}  & 0.556* & 21.0 (5) & 21.5 (3) & \cite{Czerwinski2015} & 21.9 \\\cmidrule{1-2}
        \isotope[97]{Sr}  & 0.831* & 530 (22) & 504 (8) & \cite{Rzaca2018} & 395 \\
                          & 0.308\;\; & 200 (10) & 165 (4) & \cite{Czerwinski2015} & 169 \\\cmidrule{1-2}
        \isotope[98]{Y}   & 1.181* & 740 (30) & 780 (30) & \cite{ND98} & 780 \\
                          & 0.496\;\; & 6740 (140) & 6900 (50) & \cite{ND98} & 6900 \\
                          & 0.375\;\; & 37.8 (13) & 35.2 (5) & \cite{ND98} & 35.2 \\ 
                          & 0.171\;\; & 680 (30) & 630 (20) & \cite{ND98} & 630 \\\cmidrule{1-2}
        \isotope[97]{Zr}  & 1.264* & 106 (11) & 102.8 (24) & \cite{ND97} & 102.8 \\\cmidrule{1-2}
        \isotope[99]{Zr}  & 1.039\;\; & 29 (4) & 54 (10) & \cite{Urban2003} & 54\\
                          & 0.252* & 345 (12) & 336 (5) & \cite{Boulay2020} & 293\\
                          & 0.122\;\; & 1.01 (3) & 1.08 (2) & \cite{Pfeil2023} & 1.07 \\\cmidrule{1-2}
        \isotope[101]{Zr} & 0.942* & 18.2 (19) & 16 (2) & \cite{Urban2004} & 16 \\\cmidrule{1-2}
        \isotope[108]{Tc} & 0.331* & 116 (3) & 94 (10) & \cite{Hwang2004} & -- \\
                          & 0.177\;\; & 2.81 (4) & -- & -- & -- \\\cmidrule{1-2}
        \isotope[130]{Sn} & 2.085* & 50 (5) & 63 (4) & \cite{Pietri2011} & 52 \\\cmidrule{1-2}
        \isotope[132]{Sb} & 0.254* & 112.0 (27) & 102 (4) & \cite{ND132} & 102 \\\cmidrule{1-2}
        \isotope[132]{Te} & 2.723\;\; & 3500 (300) & 3520 (90) & \cite{Kim2017} & 3700\\
                          & 1.925\;\; & 28100 (1300) & 28500 (900) & \cite{Kumar2022} & 28100 \\
                          & 1.775* & 137 (8) & 133.1 (35) & \cite{Kumar2022} & 145 \\\cmidrule{1-2}
        \isotope[133]{Te} & 1.610* & 106.5 (44) & 100 (5) & \cite{ND133} & 100 \\\cmidrule{1-2}
        \isotope[134]{Te} & 5.804\;\; & 16.7 (8) & 16.4 (17) & \cite{Hafner2021} & 18 \\
                          & 1.691* & 172.5 (16) & 164.1 (9)  & \cite{ND134} & 164.1 \\
                          & 1.576\;\; & 1.26 (2) & 1.4 (1) & \cite{Hafner2021} & 1.36 \\\cmidrule{1-2}
        \isotope[135]{Te} & 1.555* & 500 (35) & 511 (20) & \cite{ND135} & 511 \\\cmidrule{1-2}
        \isotope[136]{Xe} & 1.892* & 2950 (50) & 2920 (30) & \cite{Kim2017} & 2950 \\
                          & 1.694\;\; & 1.28 (2) & 1.293 (17) & \cite{Mach1995} & 1.293 \\\cmidrule{1-2}
        \isotope[137]{Xe} & 1.935* & 7.8 (5) & 10.1 (9) & \cite{ND137} & 10.1 \\\cmidrule{1-2}
        \isotope[146]{La} & 0.620* & 14.0 (5) & 14 (1) & \cite{Wang2017} & 20
        \\\cmidrule{1-2}
        \isotope[147]{Ce} & 0.401* & 6 (2) & -- & -- & -- \\\cmidrule{1-2}
        \isotope[152]{Nd} & 2.243* & 71 (6) & 63 (7) & \cite{Yeoh2010} & 63 \\
        \bottomrule
    \end{tabular}
\end{table*}

\begin{figure}[ht]
    \centering
    \includegraphics[width=0.99\linewidth]{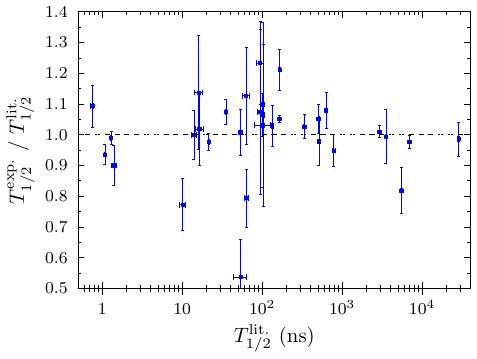}
    \caption{Ratios of half-lives measured in this work over literature values.}
    \label{fig:ratio}
\end{figure}

The ratios between all the half-lives measured in this work and the ones extracted from the literature (see Table~\ref{tab:results}) are represented in Figure~\ref{fig:ratio}, except for the newly identified isomer in \isotope[108]{Tc} and \isotope[147]{Ce}. From such a representation, the large range of half-lives that could be accessed with the VESPA setup is well visible (5 orders of magnitude). We can observe that our results agree well with the literature values, although no perfect agreement is found. The ratios do not seem to show any dependence on the half-life, indicating that our method performs equally well over the whole time range.
A large majority of the half-lives measured in this work (25 over 32) agree within 4\% ($1\sigma$) to the literature value. For some outliers, a more important deviation is observed.
The most striking difference is the half-life of the \qty{1.039}{\MeV} isomeric state in \isotope[99]{Zr}.

In Tab.~\ref{tab:results}, half-lives extracted from the RIPL3 database~\cite{RIPL} (2023 update) are also compared to the ones measured in this work. Indeed, as simulation code may rely on such database, local effects may arise from incorrect isomer half-life in calculations.

In the following subsections, dedicated isomers of interest will be presented and discussed. In particular, we extracted half-lives in \isotope[108]{Tc} and \isotope[147]{Ce}, that have been measured for the first time.

\subsubsection{\texorpdfstring{\isotope[108]{Tc}}{108Tc}}

\begin{figure}[ht]
    \centering
    \includegraphics[width=0.99\linewidth]{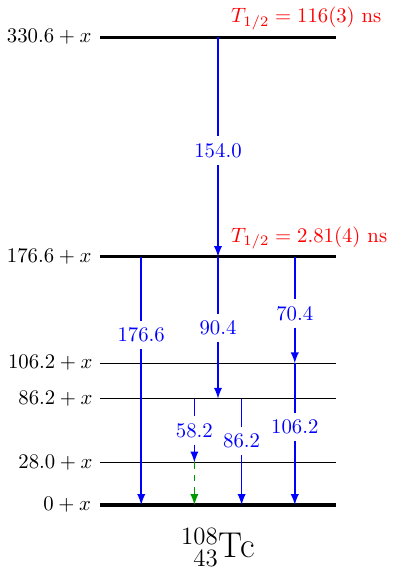}
    \caption{Level scheme of \isotope[108]{Tc} as proposed in Ref.~\cite{Hwang1998}. The half-lives of the isomeric states that have been measured in this work have been noted in red, and the $\gamma$-rays energies used in this work are given in blue.}
    \label{fig:schemeTc}
\end{figure}

Although $\gamma$-rays from the \isotope[108]{Tc} isomer were already observed in the 1970 years (see e.g. in Ref.~\cite{John1970}), no level scheme was proposed at that time.
Nevertheless, our data are consistent with the level scheme proposed in Ref.~\cite{Hwang1998} and represented in Fig.~\ref{fig:schemeTc}.
We derived from our data that the level at $E^*=(330.6+x)$~keV is a consistent candidate for the \qty{116}{\ns} isomeric state, from all $\gamma$-$\gamma$ coincidences reported in Fig.~\ref{fig:schemeTc}. 

\begin{figure}[ht]
    \centering
    \includegraphics[width=0.99\linewidth]{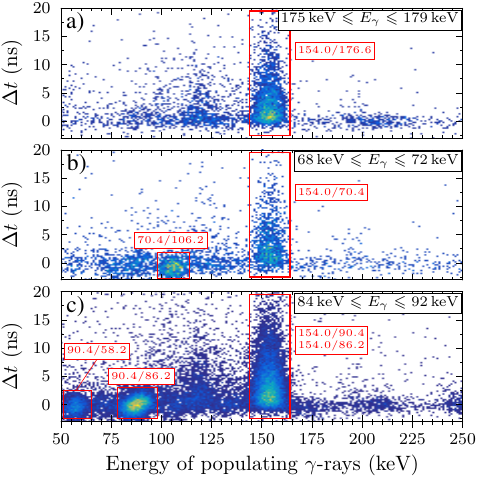}
    \caption{Populating $\gamma$-ray energy vs time difference (gate/populating) spectra obtained for different isomeric transitions in \isotope[108]{Tc}, from the multiple isomers analysis. $\gamma$-$\gamma$ cascades from \isotope[108]{Tc} level scheme are highlighted in red.}
    \label{fig:Tc_delayed}
\end{figure}

Besides, we measured for the first time the short-lived state at $E^*=(176.6+x)$~keV. From the multiple isomers analysis, we undoubtedly observed that this state is populated by a $\qty{154}{\keV}$ $\gamma$-ray, before decaying with a well-defined half-life, as represented in Figure~\ref{fig:Tc_delayed}.
In this figure, this underlying isomeric state populated by the \qty{154}{\keV} $\gamma$-ray is clearly visible, as well as other $\gamma$-ray coincidences.
These observations are consistent with the level scheme presented in Fig.~\ref{fig:schemeTc}. The half-life of this short-lived isomeric state has been measured to be $\qty{2.81}{\ns}\pm(0.02)_{\text{stat}}\pm(0.03)_{\text{syst}}$.

Additional interesting features can be noticed in Figure~\ref{fig:Tc_delayed}. In the region corresponding to the 90/86 and 86/90 coincidences from Fig~\ref{fig:Tc_delayed}c, the time distribution seems to shift with the energy between 82 and 94 keV. Indeed, in this region the anti-delayed and delayed time distributions corresponding to the \qty{90.4}{\keV} $\to$ \qty{86.2}{\keV} cascade are mixed, as the $\gamma$-rays cannot be resolved due to the energy resolution of the detectors. This could indicate the presence of a very short-lived ($\hf<\qty{1}{\ns}$) state at $E^*=(86.2+x)\,\si{\keV}$.
Finally, in Fig.~\ref{fig:Tc_delayed}a/c, a structure can be observed around $E_\gamma\simeq\qty{120}{\keV}$. This can correspond to feeding $\gamma$-rays populating the $E^*=(176.6+x)\,\si{\keV}$ state, e.g. from the  $E^*=(293.2+x)\,\si{\keV}$ and $(413.1+x)\,\si{\keV}$ states.

\subsubsection{\texorpdfstring{\isotope[147]{Ce}}{147Ce}}

A short-lived isomer was found for the $\gamma$-$\gamma$ coincidence with $\gamma$-ray energies of \qty{118}{\keV} and \qty{283}{\keV}. The average post-neutron masses of the associated light and heavy fragments are $A_L=(100.5\pm0.2)\,\mathrm{u}$ and $A_H=(147.8\pm0.2)\,\mathrm{u}$, respectively. As the mass determination is known for having slight charge-dependent deviation due to neutron emission~\cite{Piau2022}, these characteristics are consistent with the known level scheme of \isotope[147]{Ce} from the excited level $E^*=\qty{401.1}{\keV}$~\cite{ND147}. The nuclear charge distribution obtained from the IC (see Section~\ref{sec:chargeIC}) also supports this identification.
Furthermore, this isotope has already been observed in spontaneous fission of \isotope[252]{Cf}~\cite{Li2014}, although no isomeric state was reported.
This could be explained by the time resolution of HPGe detectors, which is much worse than for \labr{}, and is of particular importance for such short-lived isomer, the measured half-life being $T_{1/2}=\qty{6+-2}{\ns}$.

It is noteworthy that for short-lived isomer with a half-life below \qty{10}{\ns}, such as \isotope[147]{Ce} and \isotope[137]{Xe}, most of the data
have been cut out during the analysis, being below the $T=\qty{35}{\ns}$ threshold used to gate out $(n,n')$ events.

\subsubsection{Other noteworthy isomers}

In this final subsection, we show how the present data could be beneficial for the evaluation community, besides the new isomers previously reported.

A first interesting result is the half-life of the \isotope[97]{Sr} isomer at $E^*=\qty{831}{\keV}$.
In the present evaluated data~\cite{ND97}, an inconsistency between two measurements of this half-live is reported by the evaluator. Since then, the half-life have been re-measured a couple of times, addressing this issue~\cite{Rzaca2018,Esmaylzadeh2019}.
The result obtained in the present work with VESPA is consistent with these new measurements.

Another point of interest concerns the short-lived (\qty{0.82}{\ns}) isomer ($E^*=\qty{142}{\keV}$) in \isotope[92]{Rb}. This is, to our knowledge, the first time that a the measurement of this half-life is published, although the isomer is already known since 1972, but was only reported as private communication in the evaluated data~\cite{ND92}. 

\subsection{Charge calibration of the IC} \label{sec:chargeIC}

The identified isomers constitute a large data set of known post-neutron fission fragments, of various mass and nuclear charge (see Fig.~\ref{fig:chart}).

As initially demonstrated in the pioneer work from Ref.~\cite{Budtz1987}, a twin Frisch-grid ionization chamber could be used to estimate the nuclear charge of ﬁssion fragments, from a ratio between electron-ion pair track position in both sides of the chamber. In Ref.~\cite{Budtz1987}, the authors based their analysis on events having a high Total Kinetic Energy (TKE), corresponding to neutron-less fissions (excitation energy of the fragments below neutron emission threshold), for a better mass determination. We propose here to generalize their approach using the isomer determination previously described.
It should be noted that such a calibration presents several difficulties, due to the low kinetic energy of the fragments, that is below \qty{2}{\MeV\per u}, and the correlation between the various fission fragment characteristics (mass, nuclear charge and kinetic energy).

\begin{figure}[ht]
    \centering
    \includegraphics[width=0.99\linewidth]{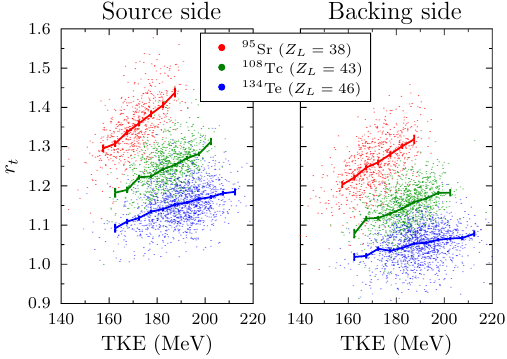}
    \caption{Evolution of the electron drift-time ratio $r_t$ as a function of TKE gated on isomers, depending on the side where the light fragments have been detected. Thick lines are the profiles associated to each distribution, i.e., $\avg{r_t\,\vert\,\tke}$.}
    \label{fig:RtTKE}
\end{figure}

The IC used in the VESPA setup can return the pulse height at each anode, to obtain the energy of both fission fragments, as well as average electron drift times ($\bar{t}$), to deduce their position along the $z$ axis~\cite{Gook2016}. We define the drift-time ratio ($r_t$) as the ratio between electron drifts time in each side of the chamber:
\begin{equation}
    r_t=\bar{t}_L / \bar{t}_H
\end{equation}
where $L$ ($H$) refers to the light (heavy) fission fragment.
This quantity is closely related to the one defined in Ref.~\cite{Budtz1987} to estimate the nuclear charge split of the fission fragments, and has also been mentioned in a more recent paper~\cite{Gaudefroy2017}. However, as illustrated in Fig.~\ref{fig:RtTKE}, this ratio depends not only on the nuclear charge of the fragments, but also on their kinetic energy.

Hence, we construct the following charge calibration function, for fission events with $\mathrm{TKE}>\qty{120}{\MeV}$:
\begin{equation}
    \hat{Z}_{L,k} = Z_\text{CN}/2 - \dfrac{r_t-\avg{r_t}_{\text{sym},k}}{a_k\times(\mathrm{TKE}-\avg{\mathrm{TKE}})+b_k}
    \label{eq:ZLofRT}
\end{equation}
where $Z_\text{CN}/2$ and $\avg{r_t}_\text{sym}$ are the nuclear charge and mean drift-time ratio of the fragments when the nucleus undergoes symmetric fission : $A_L=A_H=126$ in the case of \isotope[252]{Cf}(sf). It is generally admitted that for such fragmentation both fission fragments have the same nuclear charge, that is $Z_L=Z_H=Z_\text{CN}/2$~\cite{Wahl1988}. $\avg{\mathrm{TKE}}$ is the average post-neutron TKE, that is \qty{181.6(7)}{\MeV} from our measurement. The $k$ index is used to designate the chamber side where the light fragments is emitted (source or backing).

\begin{figure}[ht]
    \centering
    \includegraphics[width=0.99\linewidth]{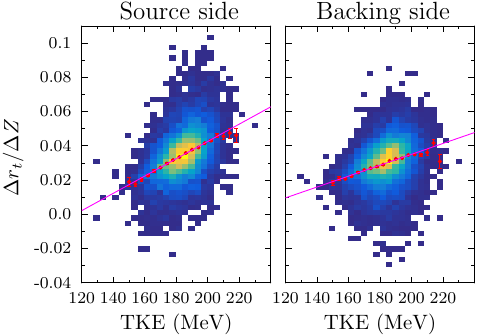}
    \caption{Distribution of $(r_t-\avg{r_t}_\text{sym}) / (Z_L - Z_\text{CN}/2)$ vs TKE in both chamber sides, from \isotope[95][38]{Sr}, \isotope[108][43]{Tc}, \isotope[134][52]{Te} and \isotope[136][54]{Xe}. Red points correspond to the profile along TKE of these distributions: $\avg{\Delta r_t / \Delta Z\;\vert\;\tke}$, while the fit result is represented by a magenta line.}
    \label{fig:calibZL}
\end{figure}

The $\avg{r_t}_\text{sym}$ parameters of the calibration has been obtained from symmetric fission, that is : $A_H-A_L\leqslant{}\qty{0.1}{u}$.
To get the remaining $a$ and $b$ parameters of the calibration function, four isomers with high yields were used: \isotope[95][38]{Sr}, \isotope[108][43]{Tc}, \isotope[134][52]{Te} and \isotope[136][54]{Xe}. These isomers were selected by gating on $\gamma$-$\gamma$ coincidences. 
The resulting $(r_t-\avg{r_t}_\text{sym}) / (Z_L - Z_\text{CN}/2)$ distributions are shown in Figure~\ref{fig:calibZL}, together with the associated linear functions $f(\mathrm{TKE})=a\times(\mathrm{TKE}-\avg{\mathrm{TKE}})+b$ used to fit them. Corresponding calibration parameters $a$ and $b$ extracted from these fits are given in Table~\ref{tab:rtCalib}.

\begin{table}[h]
    \renewcommand{\arraystretch}{1.2}
    \centering
    \caption{Parameters of the charge calibration of the VESPA IC, for each side of the chamber.}
    \begin{tabular}{l@{\hspace*{2em}}c@{\hspace*{2em}}c}
    \toprule
    & Source side & Backing side \\\midrule
    $\avg{r_t}_\text{sym}$ & 1.041 & 0.962 \\
    $a\times10^4$MeV & \num{5.05(11)} & \num{3.18(13)} \\
    $b\times10^2$ & \num{3.33(1)} & \num{2.92(1)} \\
    \bottomrule
    \end{tabular}
    \label{tab:rtCalib}
\end{table}

\begin{figure}[ht]
    \centering
    \includegraphics[width=0.99\linewidth]{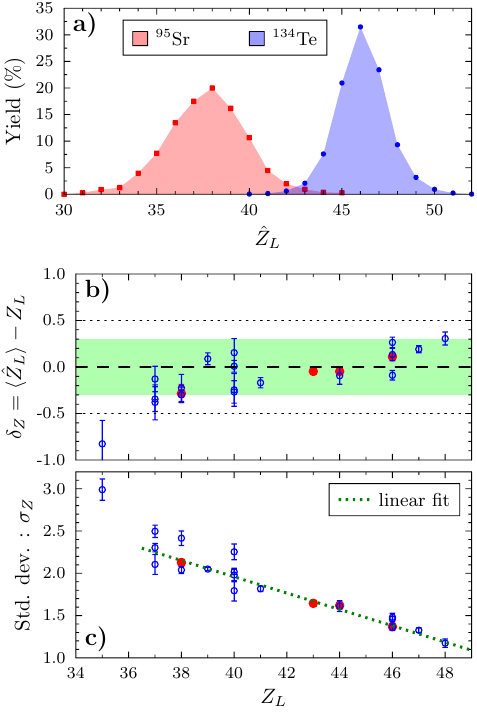}
    \caption{Results of the charge calibration of the IC. a) Measured $Z_L$ distributions of \isotope[95][38]{Sr} and \isotope[134][52]{Te} isomers (normalized). Note that the light partner of Te ($Z_H=52$) is Pd ($Z_L=46$) in the spontaneous fission of \isotope[252]{Cf} ($Z_\text{CN}=98$). b) Validation of the charge calibration using the isomers: the average measured nuclear charge $\avg{\hat{Z}_L}$ is compared to the true $Z_L$. The four isomers used in the calibration procedure are depicted as red dots. The $\vert\delta_Z\vert\leqslant{}0.3$ area is colored in green. c) Standard deviation of the measured $\hat{Z}_L$ distributions as a function of the nuclear charge.}
    \label{fig:ZLfromIC}
\end{figure}

Using the calibration parameters listed in Table~\ref{tab:rtCalib}, the nuclear charge $\hat{Z}_L$, associated to the fission fragments, could be extracted from the VESPA IC. This charge cali\-bration has been validated using the isomers identified in this work, as represented in Figure~\ref{fig:ZLfromIC}. From these figures, we can notice that this calibration procedure yields satisfactory results: the average measured nuclear charge remains in agreement with the true one ($Z_L\pm0.5$) between the calibration points (Fig~\ref{fig:ZLfromIC}b). However, we observe that extrapolation of the calibration below $Z_L=38$ should be limited. Although the results remain valid for $Z_L=37$, they become incorrect for lighter charge ($Z_L=35$). For $Z_L\geqslant{}37$, we can define a systematic uncertainty on $Z$ amounting to $\pm0.3$, as represented in Figure~\ref{fig:ZLfromIC}b as a green area.
It can also be observed from Figures~\ref{fig:ZLfromIC}a and~\ref{fig:ZLfromIC}c that the width of the obtained nuclear charge distributions increases as the fragmentation becomes more asymmetric (i.e., $Z_L$ becomes smaller).
This effect is due to the form of Eq.~\eqref{eq:ZLofRT}: the further away the fragments are from the symmetry (i.e., the larger $r_t-\avg{r_t}_\text{sym}$), the more sensitive $\hat{Z}_L$ is to fluctuations in TKE. For the same reason, the fragments far from symmetry are more sensitive to the uncertainties on the calibration parameters from Tab.~\ref{tab:rtCalib} than the ones close to symmetry.
The FWHM of the measured $Z_L$ distributions between 37 and 49 takes values between 5.9 and 2.5. These results are comparable to the ones mentioned in Ref.~\cite{Gaudefroy2017}.

\begin{figure}
    \centering
    \includegraphics[width=0.99\linewidth]{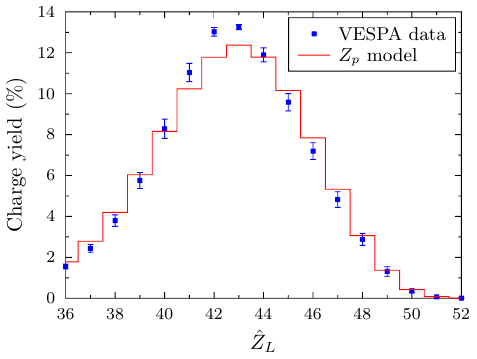}
    \caption{Measured light nuclear charge yield obtained with VESPA IC after calibration (blue). Uncertainties correspond to charge deviation within [-0.3,0.3] (see text). The red histogram corresponds to Wahl's $Z_p$ model, folded with the IC charge resolution.}
    \label{fig:YZvespa}
\end{figure}

From this charge calibration, the nuclear charge yield of light fragments from \isotope[252]{Cf} spontaneous fission can be obtained from the VESPA data, as depicted in Figure~\ref{fig:YZvespa}. The uncertainties are mostly systematic, corresponding to random fluctuation of the charge, uniformly distributed in the range [-0.3,0.3], to take into account the systematic deviation observed for isomers (Fig.~\ref{fig:ZLfromIC}b). The average measured nuclear charge is then $\avg{\hat{Z}_L}=43.42\pm0.17$. Note that Fig.~\ref{fig:YZvespa} represents the raw nuclear charge yield as measured with the IC. Unfolding this distribution with an accurate response function obtained from the isomers data (Fig.~\ref{fig:ZLfromIC}) would lead to a more physical distribution, but this complex analysis would be the subject of a forthcoming article. This distribution can still be compared to the $Z_p$ model introduced by Wahl~\cite{Wahl1988}, as represented in Fig.~\ref{fig:YZvespa}. This nuclear charge distribution has been obtained from the $Z_p$ model using the pre-neutron mass distribution $Y(A)$ measured in this work, and has been folded by the charge resolution of the IC obtained from Fig.~\ref{fig:ZLfromIC}c.

\section{Conclusion} \label{sec:conclu}

This paper has presented results of a thorough analysis of fission fragment isomers produced in the spontaneous fission of \isotope[252]{Cf} using the VESPA setup.
First, a systematic study of the half-lives of various isomers have been conducted, using two complementary analysis methods. First, $\gamma$-$\gamma$ coincidences of late-emission prompt fission $\gamma$-rays were used to identify isomers and estimate half-lives in simple cases. Second, a multiple isomers analysis was carried out to disentangle isomeric states populating each others. The results obtained in this work can be valuable for the data evaluation community. In particular, half-lives of the short-lived isomeric state in \isotope[108]{Tc} were measured for the first time. Also, our data indicate the existence of a short-lived (\qty{6}{ns}) isomeric state, $E^* = \qty{401}{\keV}$, that has been assigned to \isotope[147]{Ce}.
In total, 34 half-lives have been measured, from the nanosecond scale to tens of microseconds, with a careful estimate of associated uncertainties.   
This work demonstrates the great strength and versatility of the VESPA setup, made of a twin Frisch-grid ionization chamber in conjunction with fast scintillation detectors such as \labr{}. The main advantage of such setup is to be sensitive to half-lives across several orders of magnitude and to short-lived isotopes, using a rather simple analysis method.

The isomers obtained in this work have also been used to calibrate the IC against the nuclear charge of the fission fragments. Even though such measurement is made difficult by the low kinetic energy of the fragments and the mass resolution induced by the double kinetic energy method, the results are promising.

Based on all these encouraging results, a new generation of the VESPA setup is being built. This augmented setup consists in a larger array of both \labr{} and $\mathrm{CeBr}_3$ $\gamma$-ray detectors placed around a twin Frisch-grid ionization chamber, similar to the one used in this work.
Adding these new detectors improves the overall efficiency of the setup. This will lead to better statistics, and/or better selectivity, e.g. by means of triple coincidences in relevant cases.

\backmatter

\bmhead{Acknowledgements}
This work has been carried out in the framework of the SINET project funded by the CEA.

\bibliography{references}

\end{document}